\documentclass[
aps,superscriptaddress,floatfix,preprint
]{revtex4-1}

\usepackage{graphicx}
\usepackage{hyperref}
\usepackage{xcolor}
\usepackage{amsmath}
\usepackage{soul}

\title{Observation of Josephson Diode Effect in a Three-terminal Josephson Device} 

\begin{abstract}
The Andreev bound state spectra of multi-terminal Josephson junctions form an artificial band structure, which is predicted to host tunable topological phases under certain conditions. However, the number of conductance modes between the terminals of multi-terminal Josephson junction must be few in order for this spectrum to be experimentally accessible. In this work we employ a quantum point contact geometry in three-terminal Josephson devices to demonstrate independent control of conductance modes between each pair of terminals and access to the single-mode regime coexistent with the presence of superconducting coupling. These results establish a full platform on which to realize tunable Andreev bound state spectra in multi-terminal Josephson junctions.
\end{abstract}

\begin{document}
\author
{Gino V. Graziano,$^{1,\dagger}$ Mohit Gupta,$^{1,\dagger}$ Mihir Pendharkar,$^{2,3}$ Jason T. Dong$^{4}$,\\
Connor P. Dempsey,$^{2}$ Chris Palmstr{\o}m,$^{2,4,5}$ and Vlad S. Pribiag$^{1 \ast}$\\
\normalsize{$^{1}$School of Physics and Astronomy, University of Minnesota, Minneapolis, Minnesota 55455, USA}\\
\normalsize{$^{2}$Electrical and Computer Engineering, University of California Santa Barbara, Santa Barbara, California 93106, USA}\\
\normalsize{$^{3}$Materials Science and Engineering, Stanford University, Stanford, California 94305, USA}\\
\normalsize{$^{4}$ Materials Department, University of California Santa Barbara, Santa Barbara, California 93106, USA}\\
\normalsize{$^{5}$ California NanoSystems Institute, University of California Santa Barbara, Santa Barbara, California 93106, USA}\\
\normalsize{$\dagger$ These authors contributed equally: Gino V. Graziano and Mohit Gupta.}\\
\normalsize{$^\ast$To whom correspondence should be addressed; E-mail:  vpribiag@umn.edu.}
}

\title{Selective Control of Conductance Modes in Multi-terminal Josephson Junctions}

\maketitle

\section*{Introduction} 

Superconductor-semiconductor heterostructures have been studied both experimentally and theoretically over the past few decades, motivated by their potential to realize topologically protected quantum states~\cite{Doh2005, Williams2012, Gunel2012, Hart2014, Pribiag2015,  Kjaergaard2017, Deacon2017, Suominen2017, Lutchyn2018, Ghatak2018, Laroche2019, Fornieri2019, Ren2019} or gate-tunable quantum bits~\cite{Casparis2018}. Such states may have applications in fault tolerant quantum information processing~ \cite{Kitaev2001,Kitaev2003,Nayak2008, Sarma2015}. Multi-terminal Josephson junctions (MTJJs) may provide a novel platform for realizing higher dimensional artificial band structures formed by the Andreev bound states (ABS) present in the junction. 
In a Josephson device with $N$ superconducting terminals, the ABS spectrum depends on the $N-1$ independent phase differences between terminals, $\phi_1, \phi_2,...,\phi_{N-1}$ , which act as quasimomenta, as well as on the scattering matrix $\hat{S}$ of the interstitial junction region. Furthermore, the ABS spectra of MTJJs are predicted to host topologically protected Weyl nodes and higher-order Chern numbers~\cite{Riwar2016, Meyer2017, Xie2017, Xie2018, Xie2022}. The energy gap between different ABS bands depends on the number of conductance modes between terminals, with theoretical efforts focusing on the case of a unity or near-unity number of interterminal modes~\cite{Riwar2016,Meyer2017,Eriksson2017}. Approaching this condition necessitates the independent control of interterminal conductance modes in a MTJJ.

MTJJs may also find application as circuit elements for coupling multiple qubits~\cite{Heck2012,Heck2014, Casparis2018, Qi2018}. Additionally, they have shown rich transport features such as the coexistence of superconducting and dissipative currents~\cite{Draelos2018}, multi-terminal fractional Shapiro steps~\cite{,Deb2018,arnault2021multiterminal}, generalizations of multiple Andreev reflections (MAR)~\cite{Nowak2019,Pankratova2020}, multi-loop superconducting interferometry\cite{Strambini2016,Vischi2017} and exotic Cooper quartet transport~\cite{Freyn2011,Pfeffer2014,Cohen2018,Huang2022}.

Previous experiments on MTJJs~\cite{Draelos2018,Pankratova2020,Gino2020} have discussed the current-space differential resistance maps in three- and four-terminal devices and its dependence on parameters such as magnetic field and a single global gate voltage, but in the regime of many conductance modes. Theoretical proposals for topological ABS spectra outline the need for a small central scattering region through which the superconducting terminals are coupled in the regime of few quantum modes, however a global gate is not ideal for implementing this experimentally. Rather, a split-gate quantum-point-contact-like design where the junction legs can be independently depleted is necessary for the transport to be localized in a central common region (Figure \ref{sem_dev}\textbf{a}).

In this work, we utilize a split-gate quantum point contact (QPC) geometry which allows selective gating of each leg of a Y-shaped three-terminal junction. With this approach, we demonstrate control over conductance modes between pairs of terminals, along with access to the single-mode regime in the junction, coexisting with superconductivity. This establishes a potential \st{new} platform for exploration of the tunable ABS spectra of MTJJ devices. We present detailed results from two device designs with different junction dimensions and different split-gate geometries. 

\section*{Results}
\subsection*{Device Architecture}

The devices are fabricated on InAs quantum well heterostructures featuring a two-dimensional electron gas (2DEG) proximitized by an epitaxial aluminum layer. High interface transparency between Al and InAs (leading to induced gap comparable to the bulk gap of Al) and coherent ballistic transport in this heterostructure have been demonstrated~\cite{Shabani2016,Kjaergaard2017, Lee2019} making it an ideal platform to realize MTJJs.  The heterostructure was grown on a semi-insulating InP(001) substrate using molecular-beam epitaxy. From the bottom, the heterostructure consists of a graded buffer of $\rm{In_xAl_{1-x}As}$ with x ranging from 0.52 to 0.81, 25 nm $\rm{In_{0.75}Ga_{0.25}As}$ super-lattice, 10.72 nm $\rm{In_{0.75}Ga_{0.25}As}$ bottom barrier, 4.54 nm InAs quantum well, 10.72 nm $\rm{In_{0.75}Ga_{0.25}As}$ top barrier. Finally, there is a 10 nm layer of epitaxial aluminum deposited on the surface of the sample. The carrier concentration and mobility of the InAs 2DEG were measured using a Hall bar geometry and found to be $n=1.22\times 10^{12}$ cm$^2$ and $\mu=9920$ cm$^2$ V$^{-1}$ s$^{-1}$ in the absence of gating (see Supplementary Fig. 1), resulting in a mean free path of of $\ell \sim 180$ nm.

\begin{figure*}
    \centering
    \includegraphics[width=1\textwidth]{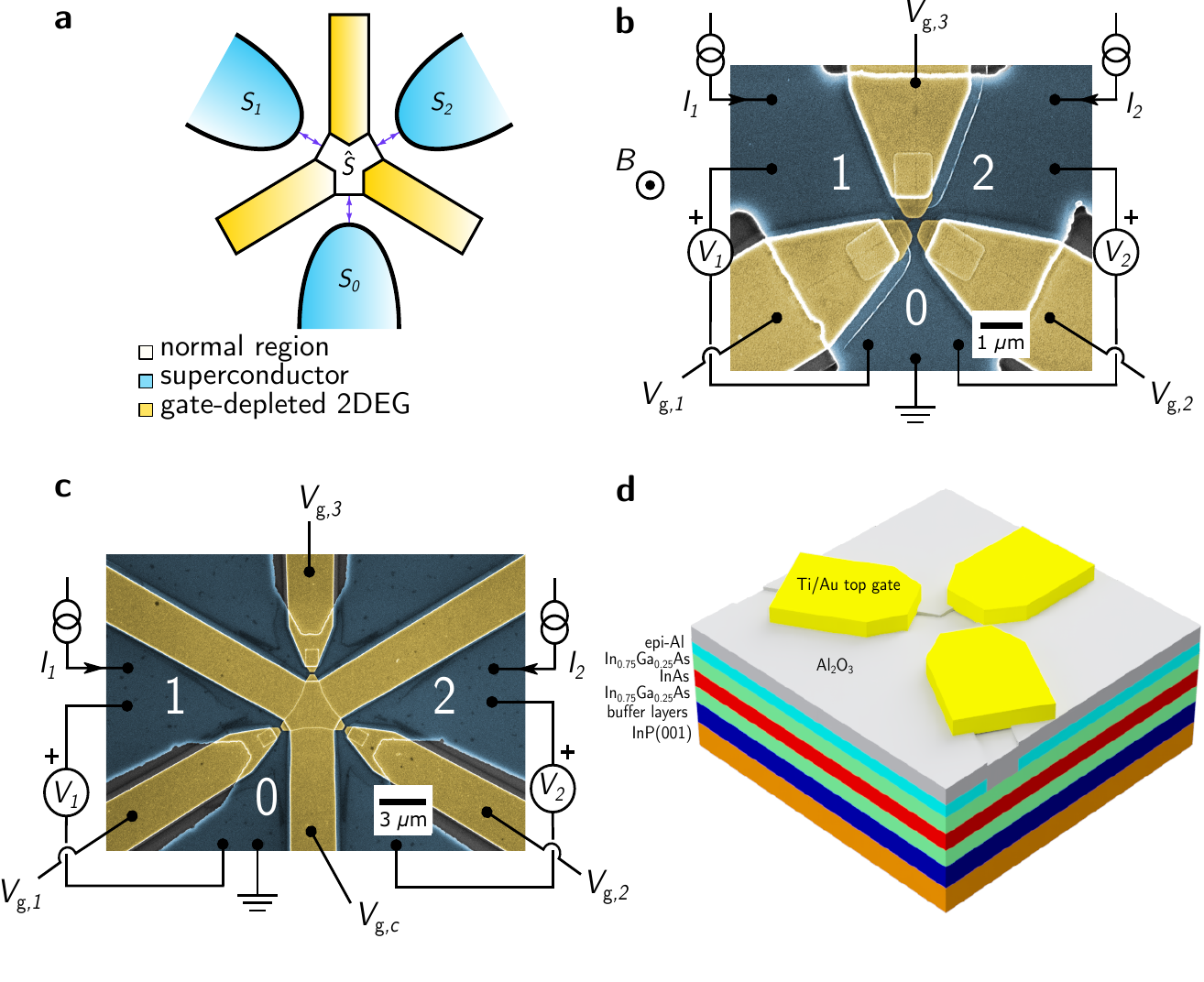}
    \caption{\textbf{Device geometry.} \textbf{a} Schematic depiction of transport in Device 1 and Device 3. The junction area under the gates, shown in yellow, can be fully depleted of carriers, leaving a central scattering region supporting a few conductance modes connected to the superconducting contacts. \textbf{b} False-color scanning electron microscope (SEM) image of Device 1, a three-terminal Josephson junction with individually tunable QPC gates, showing measurement schematic. The etched junction area is visible as the dark lines under the gates (gold-colored). \textbf{c} SEM image of Device 2, which has a central top gate that can be used both to form QPCs and to gate the central scattering area of the three-terminal junction. \textbf{d} 3D schematic of Devices 1 and 3 showing layered heterostructure.}
    \label{sem_dev}
\end{figure*}

The Y-shaped three terminal devices presented in this work have different junction widths and different split gate geometries.
Device 1 has a nominal contact spacing between superconducting electrodes of 50 nm, with three split gates as shown in Figure ~\ref{sem_dev}\textbf{b}. These split gates can deplete the 2DEG underneath forming a few-mode central region coupling each superconducting terminal. Device 2 has a nominal contact spacing of 200 nm, three split gates forming QPC-like constrictions, and also has a central top gate for independent gate control of the central scattering region  (Figure ~\ref{sem_dev}\textbf{c}). Device 3 is similar in shape to Device 1, but with electrode spacing of $\sim120$ nm. We begin by discussing transport properties of Device 1 and Device 2 and demonstrate selective gate tunability of Device 2. We then show the accessibility of single mode regime coexistent with superconductivity in these devices.

\subsection*{Transport Properties}

\begin{figure*}
\includegraphics[width=1.0\textwidth]{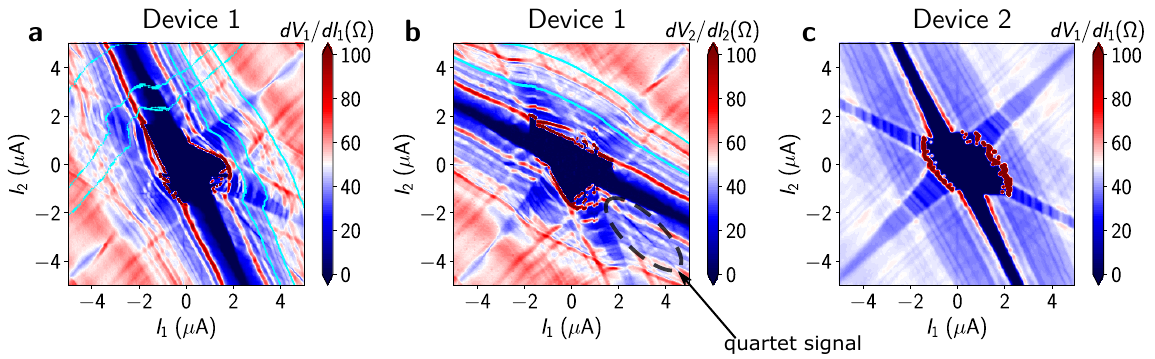} 
    \caption{\textbf{Three-terminal differential resistance maps.} \textbf{a} Measurement of the differential resistance $dV_1/dI_1$ on Device 1 with a small perpendicular magnetic field,  MAR along the lines $V_1=2\Delta/n$ is highlighted by cyan lines and along $V_1-V_2=2\Delta/n$ is highlighted by dashed cyan lines. \textbf{b} Measurement of $dV_2/dI_2$ on Device 1 with small perpendicular field, MAR along $V_2=2\Delta/n$ is shown by cyan lines. The lower resistance feature along $V_1=-V_2$ is shown by dashed ellipse which can be attributed to Cooper quartet transport. Here $I_1$ is stepped and $I_2$ is swept. \textbf{c} Measurement of $dV_1/dI_1$ on Device 2 at magnetic field $B=0$, showing MAR resonances.}
    \label{iv_char}
\end{figure*}

We perform DC current-bias measurements in a dilution refrigerator on all three devices using the configuration shown in Figure ~\ref{sem_dev}\textbf{b} and \ref{sem_dev}\textbf{c}. The superconducting data for Device 1 and Device 3 were taken at fridge temperature $T \sim 40$ mK, and Device 2 at temperature $T \sim 90$ mK. We independently control the current inputs into the epitaxial aluminum terminals 1 ($I_1$) and 2 ($I_2$) while terminal 0 is grounded. We simultaneously measure the voltages of terminals 1 ($V_1$) and 2 ($V_2$) relative to terminal 0. In a typical measurement, we step $I_2$ from negative to positive, and sweep $I_1$ from negative to positive at each value of $I_2$. We then calculate differential resistances $dV_1/dI_1$ and $dV_2/dI_2$ by discrete differentiation. The differential resistance maps show a central superconducting region where both $V_1$ and $V_2$ (Figure ~\ref{iv_char} \textbf{a, b, c}) are zero. Beyond this central region, superconducting arms are also observed approximately along $I_2=-2I_1$ (Figure ~\ref{iv_char}\textbf{a}) where only $V_1$ is zero and $I_1=-2I_2$ (Figure ~\ref{iv_char}\textbf{b}) where only $V_2$ is zero. A third superconducting arm is observed approximately along $I_1=I_2$. This feature is due to super-current being present between terminals 1 and 2 (Figure ~\ref{iv_char} \textbf{a, b, c}), while the other two arms have a nonzero resistance. The slopes of these superconducting arms in the $I_1, I_2$-plane can be understood by a resistor network model (see Supplementary Information).

The differential resistance maps exhibit rich MAR patterns. We can observe MAR as features of lower resistance along the lines $V_1= 2\Delta/n$ and $V_1-V_2=2\Delta/n$, where $n$ is an integer and $\Delta\sim 145 \mu$V is the induced superconducting gap (estimated by fitting MAR at $V_1= 2\Delta$). Figure ~\ref{iv_char}\textbf{a} shows these MAR lines highlighted in cyan for $n=2,4,6$. In Figure ~\ref{iv_char}\textbf{b} we highlight MAR along $V_2=2\Delta/n$ for $n=2,4$ in the differential resistance $dV_2/dI_2$. These three sets of MAR signatures can be understood as independent Andreev reflections between all three pairs of terminals. We also observe a signature of Cooper quartet transport~\cite{Freyn2011,Pfeffer2014,Cohen2018,Huang2022,Arnault2022}
indicated by a lower resistance feature along the line $V_1=-V_2$. The differential resistance maps can also be plotted as a function of $V_1$, $V_2$ where the quartet signature is clearly visible along the $V_1=-V_2$ diagonal (see Supplementary Fig. 2). This places Device 1 in the phase-coherent quasiballistic regime and opens up interesting possibilities for investigating cross-terminal quantum correlations.

These features are also observed for Device 2 as shown in Figure ~\ref{iv_char}\textbf{c}, despite the junction width being nearly four times larger. This is possible due to the highly transparent interface between the epitaxial aluminum and InAs quantum well of the heterostructure, and displays the robustness of our fabrication process for MTJJs and the high degree of reproducibility.
The central superconducting region is not current-symmetric in the differential resistance maps for Device 1. This indicates the presence of a small residual magnetic field resulting in asymmetric critical current~\cite{Pankratova2020,Gupta2022}, as verified in Device 2 by correcting for this residual field in our external superconducting magnet. We can observe the disappearance of this asymmetry when the perpendicular magnetic field, $B$, vanishes, as shown in Figure ~\ref{iv_char}\textbf{c}.

\subsection*{Selective Control of Conductance in Three-terminal Josephson Junctions}
A distinctive feature of these devices are their independent split top gates, enabling individual control of each leg of the Y-shaped junction. In order to demonstrate local control of the Josephson junctions formed between each pair of terminals, we can examine the results of gate depleting carriers in each of the legs selectively. Negative voltage gating of a leg results in narrowing of the width of the superconducting arm associated with it in the plot of differential resistance map. Additionally, the slopes of the lines in the $I_1$,$I_2$-plane about which superconducting features are centered change. This is due to an increase in the normal state resistance ($R_\textrm{n}$) of the leg, which affects the division of dissipative currents between the three terminals. 
When the normal state resistances in the resistor network are $R_{\textrm{n},1}, R_{\textrm{n},2}, R_{\textrm{n},3}$, the feature due to supercurrent between terminals 1 and 0 ($V_1 = 0$) falls along the line $I_2 = -\left(R_{\textrm{n},3}/R_{\textrm{n},2}+1\right) I_1$. For supercurrent between terminals 2 and 0 ($V_2=0$), this relation is $I_2 = -\left(R_{\textrm{n},3}/R_{\textrm{n},1}+1\right)^{-1} I_1$ and between terminals 1 and 2 ($V_1-V_2=0$) it lies along $I_2 = (R_{\textrm{n},1}/R_{\textrm{n},2})I_1$. Thus, we can demonstrate truly selective gating in our devices by examining the narrowing of superconducting features and their modified slopes.

\begin{figure}
    \centering
    \includegraphics[width=1.0\textwidth]{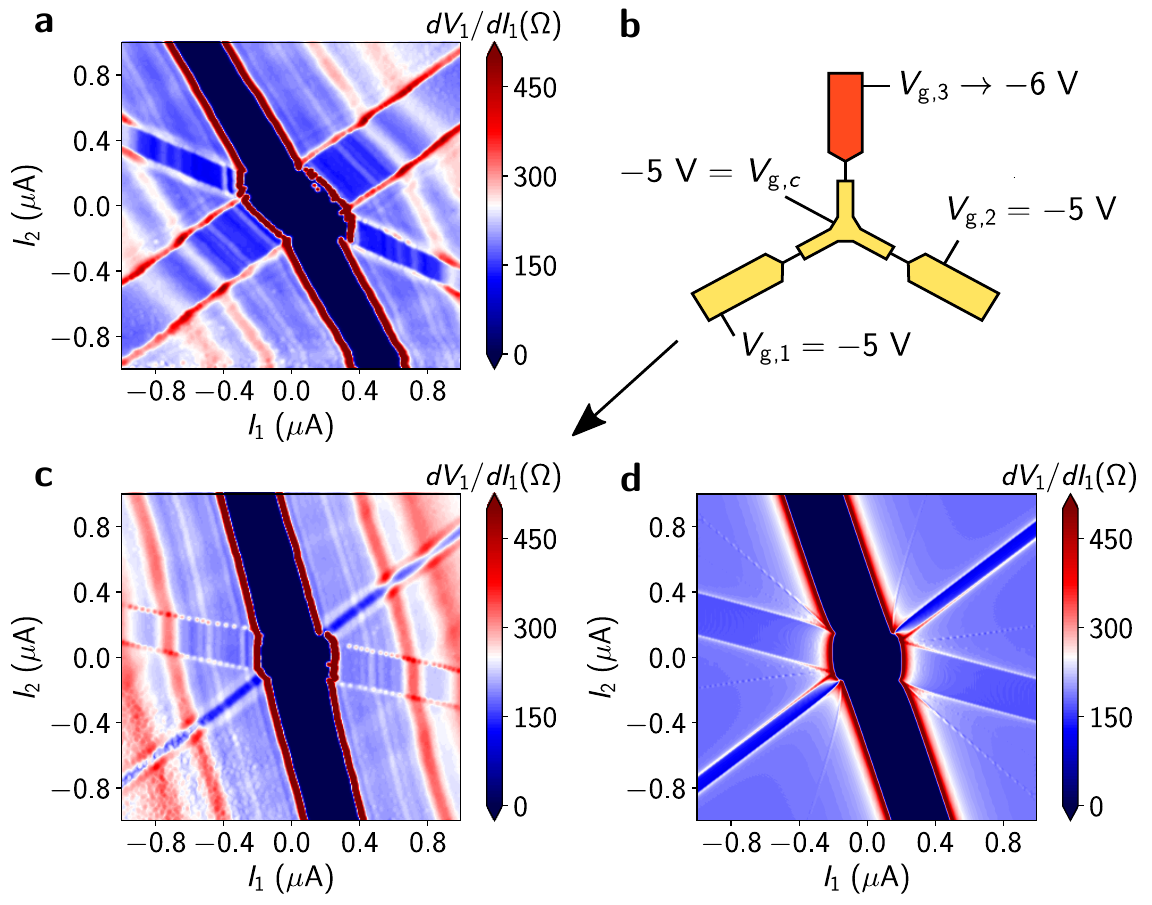}
    \caption{\textbf{Selective gating.} \textbf{a} Measurement of $dV_1/dI_1$ on Device 2, $B=0$, $T \sim 90$ mK with $V_{\textrm{g},c},V_{\textrm{g},1},V_{\textrm{g},2},V_{\textrm{g},3}=-5$ V. \textbf{b} Schematic of gate configuration for the selective gating of the junction leg between terminals 1 and 2. \textbf{c} Measurement of $dV_1/dI_1$ at $V_{\textrm{g},c},V_{\textrm{g},1},V_{\textrm{g},2}= -5$ V and $V_{\textrm{g},2}=-6$ V.  \textbf{d} RCSJ simulation of $dV_1/dI_1$ with parameters tuned to reproduce the features of experimental data.}
    \label{pref_gate}
\end{figure}

As a starting point, we measure the differential resistance $dV_1/dI_1$ with the same applied voltage on all four independent gates in Device 2 (three gates on the legs and one central gate) with $V_\textrm{g} = -5$ V (Figure \ref{pref_gate}\textbf{a}). This voltage is applied to amplify the effect of selective gating, since the superconducting features become more sensitive to gating at sufficiently negative gate voltages.  Although the applied voltage is the same, we can see a minor asymmetry of features compared to the plot at zero gate (Figure \ref{iv_char}\textbf{c}). We then decrease the voltage further only on the gate between terminals 1 and 2 ($V_{\textrm{g},3}$) (Figure \ref{pref_gate}\textbf{b}). This results in a distinct change in the differential resistance map as can be seen in Figure ~\ref{pref_gate}\textbf{c}. The superconducting arm due to supercurrent between terminals 1 and 2 ($V_1 - V_2 = 0$) dramatically decreases in width. The slope of the superconducting arm where $V_1 = 0$ has tilted toward the line $I_1 = 0$, and the $V_2 = 0$ superconducting arm has tilted toward the line $I_2 = 0$. The slope of the narrowing arm ($V_1 - V_2 = 0$) has remained unchanged. This is consistent with the limiting cases of the equations in the previous paragraph for $R_{\textrm{n},3} \gg R_{\textrm{n},1}, R_{\textrm{n},2}$. We have studied the effect of selective gating on the other two legs as well, and the slope changes were also found to be consistent with this resistor network model (see Supplementary Information). 

Additionally, we performed simulations of the system using a three-terminal resistively and  capacitively shunted junction (RCSJ) network model by solving coupled differential equations obtained by multiterminal generalization of the RCSJ model \cite{arnault2021multiterminal}. Details of the simulation and model parameters can be found in the supplementary information. This network model consists of three nodes with RCSJs between each of them, and thus contains nine independent parameters, namely the critical currents $I_{\textrm{c},i}$, normal state resistances $R_{\textrm{n},i}$ and capacitances $C_i$. The effect of gating was modeled by increasing the normal state resistance as well as decreasing the critical current of the RCSJ between two nodes relative to the others. Tuning the resistance and critical current parameters allows us to precisely reproduce the features seen in current-biased differential resistance data for preferential gating along each of the three legs (Figure \ref{pref_gate}\textbf{d}). The striped MAR features are not captured, as this is a quantum phenomenon not captured by the semiclassical RCSJ model. This conclusively shows independent control of conductance modes in each leg of the MTJJ.

\subsection*{Few Mode Three-terminal Josephson Junction}
To demonstrate tunability of conductance modes in our devices, we perform DC voltage-biased measurements on Device 1. We apply a DC source-drain voltage bias $V_{\rm{sd}}$ between a pair of terminals with the third terminal electrically floating and measure the resultant DC current $I_{\rm{meas}}$. The voltage drop across the device $V_{\rm{meas}}$ is also monitored simultaneously to exclude the effect of series resistance. We can then compute the differential resistance $d V_{\rm{meas}}/d I_{\rm{meas}}$ and differential conductance $G=d I_{\rm{meas}}/d V_{\rm{meas}}$ by discrete differentiation.

\begin{figure}
    \centering
    \includegraphics[width=1\textwidth]{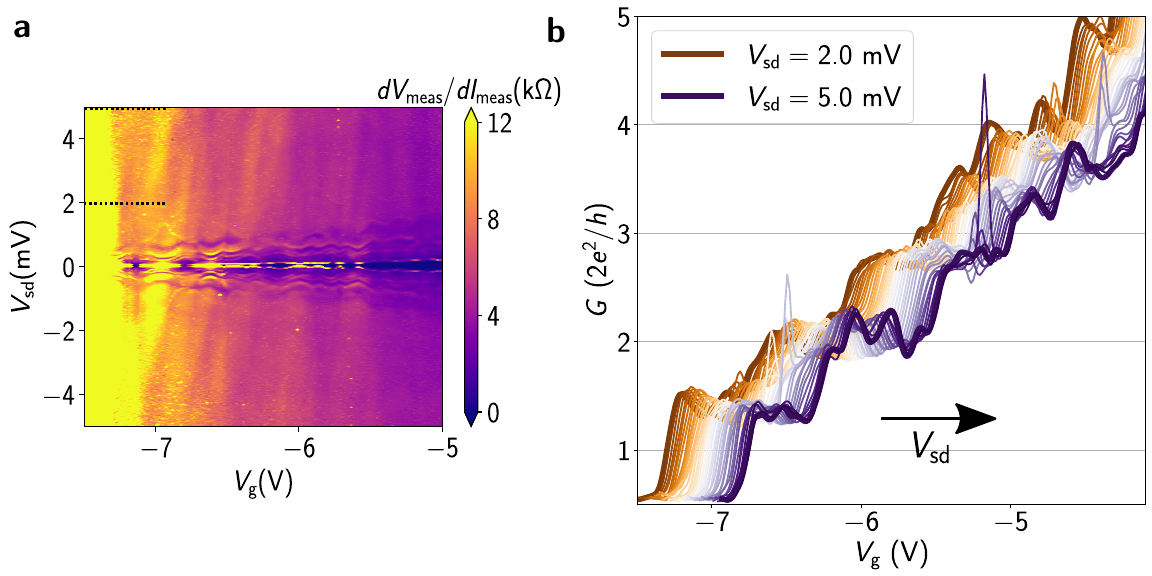}
        \caption{\textbf{Single mode MTJJ.} \textbf{a} Differential conductance as a function of source-drain bias $V_{\rm{sd}}$ and gate voltage $V_\textrm{g}$ for terminal pair 1 and 2 for Device 1 at $B=0$ and $T \sim 40$ mK. \textbf{b} Differential conductance as a function of gate voltage for different $V_{\rm{sd}}$ for Device 1 at $B=0$ and $T \sim 40$ mK. The curves correspond to $V_{\rm{sd}}$ values between 2.0 mV and 5.0 mV (shown by dashed black lines in panel (a) in increments of 0.125 mV, and are each offset along the $V_\textrm{g}$ axis (arrow indicating direction of increasing $V_{\rm{sd}}$) by 3 mV for clarity. }

    \label{qpc_data}
\end{figure}

Figure 4\textbf{a} shows a map of differential resistance between terminals 1 and 2 in Device 1 as a function of $V_{\rm{sd}}$ and gate voltage applied to all three split gates, $V_\textrm{g}$. The critical current countours are observed as areas of zero resistance and MAR is observed as areas of reduced resistance for $V_{\rm{sd}} \lesssim 2$ mV. These features show periodic oscillations as a function of gate voltage. These oscillations indicate Fabry-P\'{e}rot interference, which has been observed before in a two-terminal graphene Josephson junction~\cite{Calado2015}. This results from interference of supercurrent trajectories that travel ballistically from one contact to the other, conclusively showing ballistic transport between the two terminals. Supercurrent is present between the two measured terminals at the conductance values $\sim 1.0 \, G_0$, where $G_0 = 2e^2/h$ is the conductance quantum. We also observe conductance plateaus as a function of $V_\textrm{g}$ (Figure~\ref{qpc_data}\textbf{b}). The step height of these plateaus differs from the conductance quantum $G_0$, and the quantization weakens for higher values of conductance. This is likely due to the effect of finite source-drain bias on the conductance. At finite bias, the value of the conductance steps is determined by the number of quasi-1D subbands falling within the bias window set by $V_{\rm{sd}}$ \cite{Patel1991,Gallagher2014,Lee2019}. Such finite-bias conductance measurements is necessary due to the presence of superconductivity and MAR resonances below $V_{\rm{meas}} < 2\Delta$ \cite{Gallagher2014,Thierschmann2018,Mikheev2021}.

\begin{figure}
    \centering
    \includegraphics[width=1\textwidth]{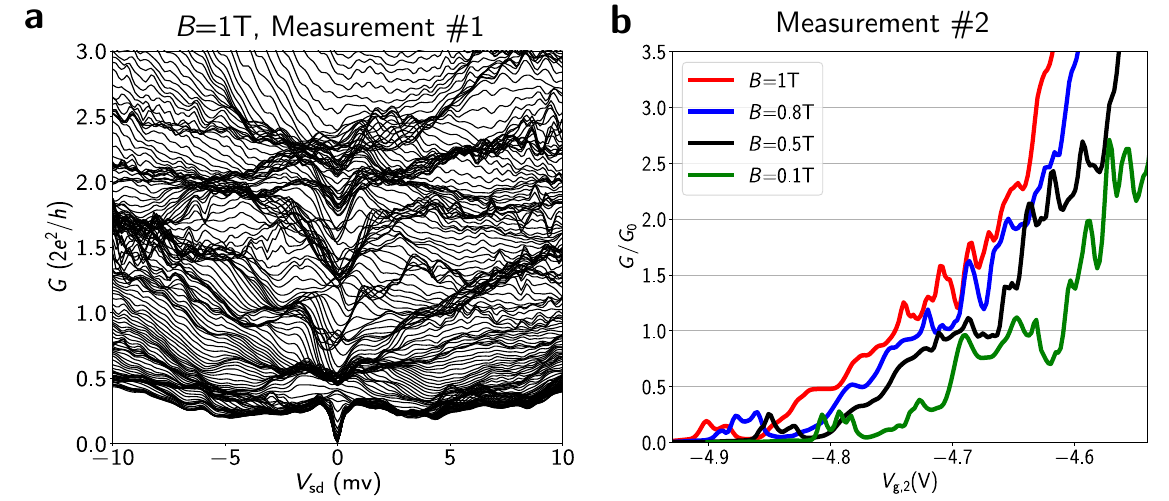}
        \caption{\textbf{Zero-bias conductance measurements.} \textbf{a} Waterfall plot of differential conductance as a function of $V_{\rm{sd}}$ for a range of gate voltages from $V_{\textrm{g},2}=-5.26$ V to $V_{\textrm{g},2}=-5.5$ V with a step size $\delta V_\textrm{g} =1.3$ mV, for Device 3. \textbf{b} Zero-bias differential conductance for Device 3 for different values of out-of-plane magnetic field. For these measurements we have set $V_{\textrm{g},1}=-6$ V and $V_{\textrm{g},3}=-3$ V. The curves are offset on the gate voltage axis by 0.02 V, 0.06 V and 0.1 V for $B=0.8$ T, $B=0.5$ T, and $B=0.1$ T respectively for clarity. All measurements were taken at $T=40$ mK.}

    \label{qpc_1T}
\end{figure}

To measure conductance at zero-bias, an out-of-plane magnetic field can be applied to eliminate superconducting effects. We measure Device 3 (lithographically identical to Device 1) in this regime. Conductance measurements are performed for the terminal pair 2 and 0 using standard lock-in techniques. The waterfall plot of conductance in Figure \ref{qpc_1T}\textbf{a} shows bunching of lines at zero bias around 0.5, 1, 1.5 and 2.0 $G_0$ due to spin splitting of the subbands. At finite bias, the waterfall plot shows bunching of curves at conductance values between integer multiples of $e^2/h$ as previously observed in two-terminal superconducting QPCs with magnetic field \cite{Gallagher2014}. This can cause the step heights to differ from the conductance quantum $G_0$, as observed for Device 1 at $B=0$ T. In a separate measurement, conductance scans are performed by varying the magnetic field and keeping $V_\textrm{sd}=0$ V. As shown in the red curve in Figure \ref{qpc_1T}\textbf{b}, conductance steps are observed near the conductance values where lines bunch at $B=1$ T.  Figure \ref{qpc_1T}\textbf{b} further shows that the plateaus become more well-defined as $B$ is increased. Additionally, resonances in the conductance data are smoothed by application of magnetic field, attributed to the suppression of coherent backscattering due to the Aharanov-Bohm phase contribution. Conductance measurements at $B=0$ T are also performed for all three pairs of the terminals and are consistent with those on Device 1 (Supplementary Fig. 5). The voltage range $V_{\textrm{g},2}$ is different between Fig. ~\ref{qpc_1T}\textbf{a} and \textbf{b} due to gate hysteresis.

This demonstrates that transport between the two measured terminals takes place via few conductance modes. Only the first few modes are individually resolved by our measurements, which could be due to a non-ideal potential profile in the central region of the junction where the confining potential may be weakened due to screening by the Al contacts. However, to resolve the ABS spectra of MTJJs only the first few conductance modes are necessary \cite{Heck2014,Riwar2016}. Moreover, theory predicts that the quantized transconductance signatures of Weyl nodes can only be resolved in the single mode limit \cite{Eriksson2017}.

Similar data is obtained for all three pairs of terminals for Device 2 as well (Supplementary Fig. 6), and the single-mode regime is accessible, coexistent with superconductivity. However, the conductance quantization is less robust than that seen in Devices 1 and 3 (Supplementary Fig. 6). This can be attributed to the mean free path in the InAs QW ($\ell \sim 180$ nm) being comparable to the junction width of 200 nm, increasing the susceptibility to scattering relative to Devices 1 and 3. It should be noted that the maximum measured resistance saturates at $\sim 100$ k$\Omega$ for Device 1 and $\sim 10$ k$\Omega$ for Device 2. For the Device 3 data in Figure \ref{qpc_1T}, we subtract this conductance contribution ($\sim 40$ k$\Omega$) at each value of magnetic field. These residual resistances can be attributed to trivial edge modes of InAs present in the etched mesa. These surface modes do not respond to a top gate and are difficult to eliminate in InAs~\cite{Noguchi1991,Olsson1996,Mueller2017,Mittag2017}, but not expected to be detrimental to the investigation of the ABS. 

\section*{Discussion}
We demonstrate phase-coherent quasiballistic transport in three-terminal split-gated Josephson devices, with access to the single quantum mode regime independently in each leg. This is the first demonstration of accessibility of all theoretical constraints necessary to observe topologically protected states formed in ABS of MTJJs. This presents a promising alternative platform to realize topological quantum states, complementary to the much explored Majorana zero modes (MZMs). Realization of topologically protected states in the Andreev bound state spectra of MTJJs also requires fine tuning of the scattering matrix of the central region, which can in principle be achieved with geometries similar to that of Device 2. Devices with more than three-terminals can be explored on the same material platform with similar gate structure, making detection of these exotic states more likely using a range of proposed approaches.

\section*{Methods}

\subsection*{Device Fabrication}
Standard electron beam lithography (EBL) and wet etching techniques were used to fabricate a mesa and the Y-shaped junction area. Approximately 40 nm of $\rm{Al_2O_3}$ dielectric was deposited using thermal atomic layer deposition (ALD). Using EBL, split gates are defined over the junction area and electrodes are deposited using electron-beam evaporation of Ti/Au (5 nm/50 nm). In a separate lithography step thicker gold contacts (Ti/Au, 5 nm/200 nm) are made to the gate electrodes~\cite{Lee2019_2}.

\subsection*{Measurement Details}

Differential resistance maps on both devices and conductance quantization data on Device 1 were obtained by low-noise DC  transport measurements in a $^3$He/$^4$He dilution refrigerator. For the conductance quantization data on Device 2 (in Supplementary Material), standard low-frequency lock-in techniques were used with a small excitation voltage and a frequency of 19 Hz. For the AC conductance measurements the raw data is corrected by subtracting the series filter and the ammeter resistances which combine to give 6.6 k$\Omega$. Low-pass Gaussian filtering was used to smooth numerical derivatives.

\section*{Data Availability}
Source data for the figures presented in this paper are available at the following Zenodo database [\url{https://zenodo.org/record/6718253}].

\section*{Code Availability}
The data plotting code and code for the performed simulations are available at the following Zenodo database [\url{https://zenodo.org/record/6718253}].

\providecommand{\noopsort}[1]{}\providecommand{\singleletter}[1]{#1}%

\section*{Acknowledgement}
We thank Alex Levchenko and Steven Koester for helpful discussions. This work was supported primarily by the National Science Foundation under Award No. DMR-1554609. The work at UCSB was supported by the Department of Energy under Award No. DE-SC0019274. The development of the epitaxial growth process was supported by Microsoft Research. Portions of this work were conducted  in the Minnesota Nano Center, which is supported by the National Science Foundation through the National Nano Coordinated Infrastructure Network (NNCI) under Award Number ECCS-1542202. 
\section*{Author Contributions}
M.G., G.G. and V.S.P. designed the differential resistance and conductance quantization experiments. G.G. fabricated and performed the transport measurements on Device 1 and 3 at UMN. M.G. fabricated and performed the transport measurements on Devices 2 and 3 at UMN. M.G. and G.G. analysed the data on all devices under the supervision of V.S.P. M.P., J.D. and C.D. grew the heterostructure under the supervision of C. J. P. M.G. wrote the numerical program for the RCSJ model calculations. M.G., G.G. and V.S.P. co-wrote the manuscript. All authors discussed the results and provided comments on the manuscript.
\section*{Competing interests}
The authors declare no competing interests.

%%%%%%%%%%%%%%%%%%%%Figures%%%%%%%%%%%%%%%%%%%%%

\end{document}

\author
{Gino V. Graziano,$^{1,\dagger}$ Mohit Gupta,$^{1,\dagger}$ Mihir Pendharkar,$^{2,3}$ Jason T. Dong$^{4}$,\\
Connor P. Dempsey,$^{2}$ Chris Palmstr{\o}m,$^{2,4,5}$ and Vlad S. Pribiag$^{1 \ast}$\\
\normalsize{$^{1}$School of Physics and Astronomy, University of Minnesota, Minneapolis, Minnesota 55455, USA}\\
\normalsize{$^{2}$Electrical and Computer Engineering, University of California Santa Barbara, Santa Barbara, California 93106, USA}\\
\normalsize{$^{3}$Materials Science and Engineering, Stanford University, Stanford, California 94305, USA}\\
\normalsize{$^{4}$ Materials Department, University of California Santa Barbara, Santa Barbara, California 93106, USA}\\
\normalsize{$^{5}$ California NanoSystems Institute, University of California Santa Barbara, Santa Barbara, California 93106, USA}\\
\normalsize{$\dagger$ These authors contributed equally: Gino V. Graziano and Mohit Gupta.}\\
\normalsize{$^\ast$To whom correspondence should be addressed; E-mail:  vpribiag@umn.edu.}
}

\title{Supplementary Information for ``Selective Control of Conductance Modes in Multi-terminal Josephson Junctions"}

\maketitle
\section{Hall Measurements}
We have constructed a Hall bar device to measure the mean free path of the 2DEG. Shubnikov-de Hass (SdH) oscillations in the longitudinal resistance ($R_{xx}$) are observed. By taking a linear fit of the inverse values of magnetic flux densities where maxima of Sdh oscillations occur, the charge density $n$ can be evaluated as:
\begin{eqnarray}
n=\frac{ 2e}{m h}
\end{eqnarray}
\begin{figure}[ht]
    \centering
    \includegraphics[width=1\textwidth]{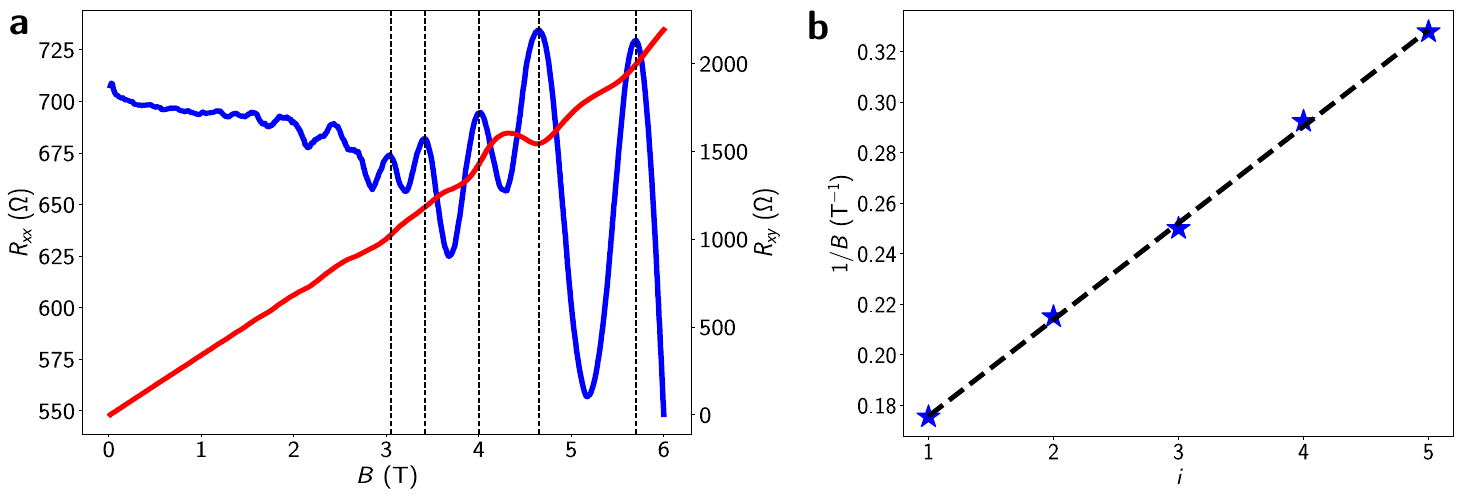}
    \caption{\textbf{a} Symmetrized longitudinal resistance $R_{xx}$ (shown in blue) and anti-symmetrized transverse resistance $R_{xy}$ (shown in red) as a function of magnetic field of the Hall bar device. To extract the period of SdH oscillations the peaks in the resistances are highlighted by dashed black lines. \textbf{b} Inverse of magnetic field values as a function of maxima of SdH oscillations (blue stars). Linear fit is shown by dashed black line. }
    \label{hall_bar}
\end{figure}
where $m$ is the slope of the linear fit. From our measurements we get $m=0.04$T$^{-1}$, this gives a charge density of $n=1.22\times 10^{12}$ cm$^{-2}$. The mobility can be evaluated by:
\begin{eqnarray}
\mu=\frac{1}{e n \rho_{xx}(B=0)}
\end{eqnarray}
$\rho_{xx}=R_{xx} (W/L)$, here $W$ is the channel width of the constructed Hall bar and $L$ is distance between he probes used to measure $R_{xx}$. From our measurements we get $\rho_{xx}(B=0)=515$ $\Omega$. This give a mobility of $\mu=9920$ cm$^2$V$^{-1}$s$^{-1}$. The mean free path, $\ell$ can be evaluated by:
\begin{eqnarray}
\ell=\frac{\hbar}{e} \mu \sqrt{ 2\pi n}
\end{eqnarray}
Using the above mentioned $\mu$ and $n$ we get $\ell\sim 180$ nm. 
\section{Differential resistance map in Voltage Space}
Differential resistance maps for Device 1 are plotted in $V_1$-$V_2$ space. The lower resistance feature along $V_1 = -V_2$ which can be attributed to Cooper quartet transport is clearly visible.
\begin{figure}[ht]
    \centering
    \includegraphics[width=1.0\textwidth]{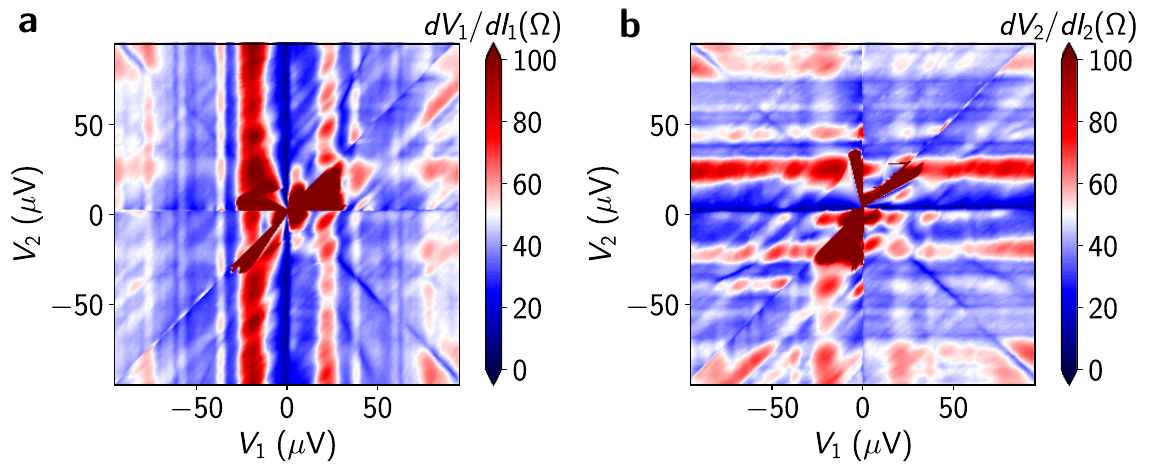}
    \caption{\textbf{a}  Measurement of $dV_1/dI_1$ on Device 1 at small magnetic field \textbf{b} Measurement of $dV_2/dI_2$ on Device 1 at small magnetic field.}
    \label{vspace_iv}
\end{figure}

\section{Resistor Network Model}

For deducing the behavior of the system under two independent current or voltage biases, we can treat the trijunction as a $\Delta$-type resistor network. The $\Delta$-type resistor network consists simply of three nodes $\{1, 2, 0 \}$ with resistors $\{ R_1, R_2, R_3\}$ connecting each pair. In the context of the real trijunction device: the nodes correspond to the three aluminum contacts, and when the system is driven into a completely resistive state, the values of the network resistances correspond to the pairwise normal state resistances $R_{\textrm{n},i}$ of the Josephson junctions between the terminals. 

\begin{figure}
    \centering
    \includegraphics[width=0.8\textwidth]{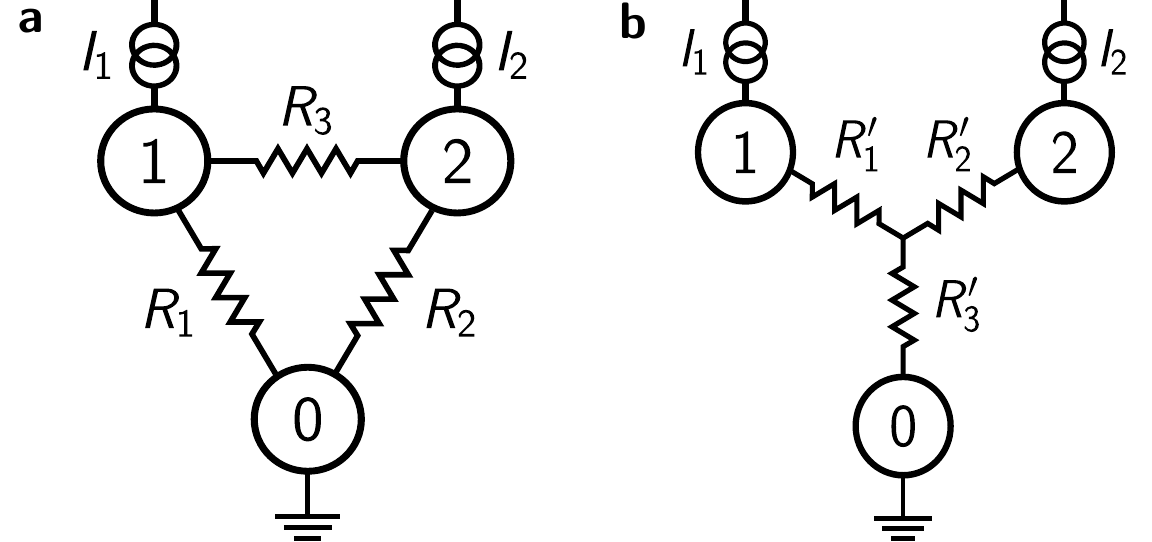}
    \caption{\textbf{a} $\Delta$ network \textbf{b} Equivalent Y network.}
    \label{fig:my_label}
\end{figure}

The transformation from $\Delta$ to Y yields the following expressions for the resistances in the Y-equivalent network:

\begin{equation}
    R_1^\prime = \frac{R_1 R_3}{R_1 + R_2 + R_3}, \,
    R_2^\prime = \frac{R_2 R_3}{R_1 + R_2 + R_3}, \,
    R_3^\prime = \frac{R_1 R_2}{R_1 + R_2 + R_3}
\end{equation}

Applying Kirchoff's laws to the Y network with node 0 grounded ($V_0 =0$) yields:

\begin{equation}
    V_1 = I_1 (R_1^\prime + R_3^\prime) + I_2 R_3^\prime, \, V_2 = I_1 R_3^\prime + I_2 (R_2^\prime + R_3^\prime), \, V_1 - V_2 = I_1 R_1^\prime - I_2 R_2^\prime
\end{equation}

Using these equations, we can set $V_1$, $V_2$ or $V_1 - V_2$ equal to 0 to identify the relations between $I_1$ and $I_2$ that yield zero voltage between terminals. The superconducting arm features of the 2D resistance map figures will be centered along these lines.

\begin{equation}
    V_1 = 0 \rightarrow I_2 = - \left( \frac{R_3}{R_2} +1 \right) I_1, \,  V_2 = 0 \rightarrow I_2 = - \left( \frac{R_3}{R_1} +1 \right)^{-1} I_1, \, V_1 - V_2 = 0 \rightarrow I_2 = \frac{R_1}{R_2}  I_1
\end{equation}

For a fully symmetric junction ($R_1,R_2,R_3 = R$) we recover that the superconducting features lie along the lines $I_2 = -2 I_1$, $I_2 = - \frac{1}{2} I_1$ and $I_1 = I_2$.

For selective gating, where one leg of the junction is under the effect of a more negative gate voltage, we can study the behavior of the superconducting feature lines by looking at the limits of equations (3), (4) and (5). Locally negatively gating one leg relative to the others will ideally increase the resistance between the two terminals that most closely span the leg.

Depleting the leg between terminals 1 and 2 increases $R_3$. Taking the limit as $R_3 \rightarrow \infty$:

\begin{equation}
    V_1 = 0 \rightarrow I_1 = 0, \, V_2 = 0 \rightarrow I_2 = 0, \, V_1 - V_2 = 0 \rightarrow I_1 = \frac{R_1}{R_2} I_1 \textrm{  (Unchanged)}
\end{equation}

Depleting the leg between terminals 1 and 0 increases $R_1$. Taking the limit as $R_1 \rightarrow \infty$:

\begin{equation}
    V_1 = 0 \rightarrow I_2 = - \left( \frac{R_3}{R_2} +1 \right) I_1 \textrm{  (Unchanged)}, \, V_2 = 0 \rightarrow I_2 = - I_1, \, V_1 - V_2 = 0 \rightarrow I_1 = 0
\end{equation}

Depleting the leg between terminals 2 and 0 increases $R_2$. Taking the limit as $R_2 \rightarrow \infty$:

\begin{equation}
    V_1 = 0 \rightarrow I_2 = - I_1, \, V_2 = 0 \rightarrow I_2 = - \left( \frac{R_3}{R_1} +1 \right)^{-1} I_1 \textrm{  (Unchanged)}, \, V_1 - V_2 = 0 \rightarrow I_2 = 0
\end{equation}

\section{RCSJ Simulation}

To model the behavior of a device with preferential gating, we simulated a three-terminal Josephson junction using an resistively and capacitatively shunted junction (RCSJ) network model. The simulation considers three nodes with three RCSJ's connecting each pair. These RCSJs between any two superconducting nodes $i,j$, have three parameters consisting of their critical current $I_{\textrm{c},ij}$, normal state resistance $R_{\textrm{n},ij}$ and capacitance $C_{ij}$. All the capacitances ($C_{i,j})$ are set to  $C = 5\times 10^{-15}$ F. This small capacitance is only necessary for the stabilization of the simulations. The current between any two nodes is then:
\begin{eqnarray}
I_{ij}= I_{\textrm{c},ij}\sin(\phi_{ij})+(\frac{1}{R_{\textrm{n},ij}})\frac{\hbar}{2e}\dot{\phi}_{ij}+C_{ij}\frac{\hbar}{2e}\ddot{\phi}_{ij}
\end{eqnarray}
Here $\phi_{ij}$ is the superconducting phase difference between $i$ and $j$ node. Due to gauge invariance we can set node 0's phase  $\phi_0=0$, this allows the notation to be simplified, with $\phi_{10}=\phi_1$, $\phi_{20}=\phi_2$ and $\phi_{12}=\phi_1-\phi_2$. We simplify also $R_{\textrm{n},10} = R_{\textrm{n},1}$, $R_{\textrm{n},20} = R_{\textrm{n},2}$ and $R_{\textrm{n},12} = R_{\textrm{n},3}$. Using Kirchhoff's law at two node 1 and 2 we get the following two equations:
\begin{eqnarray}
I_1&=&I_{\textrm{c},1} \sin(\phi_1)+(G_{\textrm{n},1}+G_{\textrm{n},3})\frac{\hbar}{2e}\dot{\phi_1}+2C\frac{\hbar}{2e}\ddot{\phi}_{1} \nonumber \\ &+&I_{\textrm{c},3}\sin(\phi_1-\phi_2)-G_{\textrm{n},3}\frac{\hbar}{2e}\dot{\phi_2}-C\frac{\hbar}{2e}\ddot{\phi}_{2}\\
I_2&=&I_{\textrm{c},2} \sin(\phi_2)+(G_{\textrm{n},2}+G_{\textrm{n},3})\frac{\hbar}{2e}\dot{\phi_2}+2C\frac{\hbar}{2e}\ddot{\phi}_{2} \nonumber \\ &-&I_{\textrm{c},3}\sin(\phi_1-\phi_2)-G_{\textrm{n},3}\frac{\hbar}{2e}\dot{\phi_1}-C\frac{\hbar}{2e}\ddot{\phi}_{1}
\label{sim_eq}
\end{eqnarray}
here $G_{\textrm{n},i}=1/R_{\textrm{n},i}$ is the conductance between any two pairs of terminals. One can rewrite Eq. \ref{sim_eq} for $\ddot{\phi_1}$ and $\ddot{\phi_2}$. This leads to a second order differential equation for the phase variable $\Phi=\begin{pmatrix} \phi_1 \\ \phi_2  \end{pmatrix}$:
\begin{eqnarray}
\frac{\hbar}{2e}\mathcal{C}\ddot{\Phi}+\frac{\hbar}{2e}\mathcal{G}\dot{\Phi}=\mathcal{I}-f(\Phi)
\end{eqnarray}
with:
\begin{eqnarray}
\mathcal{C}&=&\begin{pmatrix} 2C & -C \\ -C & 2C \end{pmatrix}\\
\mathcal{G}&=&\begin{pmatrix} G_{\textrm{n},1}+G_{\textrm{n},3} & -G_{\textrm{n},3} \\ -G_{\textrm{n},3} & G_{\textrm{n},2}+G_{\textrm{n},3} \end{pmatrix}\\
\mathcal{I}&=&\begin{pmatrix} I_1 \\ I_2 \end{pmatrix}\\
f(\Phi)&=&\begin{pmatrix} I_{\textrm{c},1} \sin(\phi_1)+I_{\textrm{c},3}\sin(\phi_1-\phi_2) \\ I_{\textrm{c},2} \sin(\phi_2)-I_{\textrm{c},3}\sin(\phi_1-\phi_2) \end{pmatrix}
\end{eqnarray}
The above second order differential equation can be solved numerically to find $\Phi(t)$ for given initial conditions and $\langle \dot\Phi\rangle$ can be calculated this allows us to get the two voltage drops $V_1=\langle\dot{\phi_1}\rangle$ and $V_2=\langle\dot{\phi_2}\rangle$ for a given value of $I_1$ and $I_2$. We utilize the approach outlined in the Ref.~\cite{Arnault2022} using Pytorch library to solve in parallel for a grid of $I_1$ and $I_2$ allowing for rapid computation. Source code for the performed simulations is provided with this manuscript.

\begin{figure}
    \centering
    \includegraphics[width=1\textwidth]{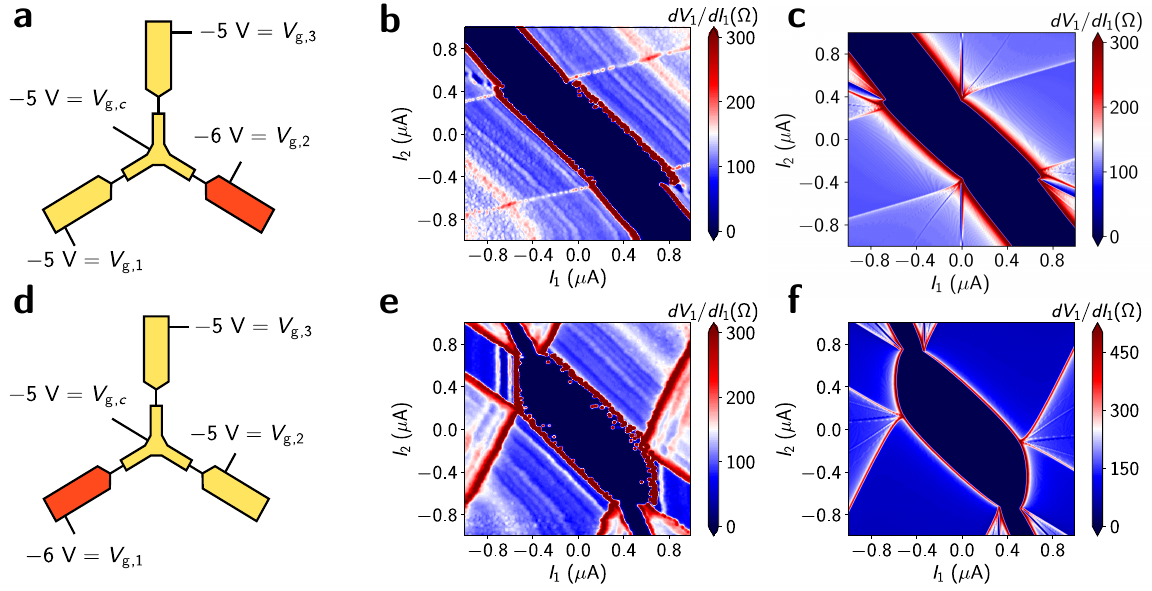}
    \caption{\textbf{a} Schematic of gate configuration for the gating of junction leg between terminal 0 and 2. \textbf{b} Measurement of $dV_1/dI_1$ at $V_{\textrm{g},c},V_{\textrm{g},1},V_{\textrm{g},3}= -5$V and $V_{\textrm{g},2}=-6$ V.  \textbf{c} RCSJ simulation of $dV_1/dI_1$ with parameters tuned to match the features of experimental data in \textbf{b}. \textbf{d} Schematic of gate configuration for the gating of junction leg between terminal 0 and 1. \textbf{e} Measurement of $dV_1/dI_1$ at $V_{\textrm{g},c},V_{\textrm{g},2},V_{\textrm{g},3}= -5$V and $V_{\textrm{g},1}=-6$ V.  \textbf{f} RCSJ simulation of $dV_1/dI_1$ with parameters tuned to match the features of experimental data in \textbf{e}.}
    \label{sgating_0102}
\end{figure}

Supplementary Fig. \ref{sgating_0102}\textbf{a} and \ref{sgating_0102}\textbf{d} shows schematics of preferential gating along the two junction legs not included in the main text. The $I_{\textrm{c},i}$ and $R_{\textrm{n},i}$ values in the simulation were tuned by iteration to match the experimental data (Supplementary Fig. \ref{sgating_0102}\textbf{b} and \textbf{e}) taken in these configurations. The values used in the presented simulation figures (Figure. 3\textbf{d} in main text and Supplementary Fig. \ref{sgating_0102}\textbf{c}, \textbf{f}) can be found in Supplementary Table \ref{t1}.

\begin{table}
\begin{tabular}{|c|c|c|c|c|c|c|c|c|c|}
    \hline
    Figure & $I_{\textrm{c},1}$ & $I_{\textrm{c},2}$ & $I_{\textrm{c},3}$ & $R_{\textrm{n},1}$ & $R_{\textrm{n},2}$ & $R_{\textrm{n},3}$ \\
    \hline
   Figure 3\textbf{d} (main text) & 180 nA & 130 nA & 30 nA & 220 $\Omega$ & 280 $\Omega$ & 600 $\Omega$ \\
    \hline
    Supplementary Fig. 4\textbf{c} & 350 nA & 50 nA & 350 nA & 125 $\Omega$ & 425 $\Omega$ & 125 $\Omega$ \\
    \hline
    Supplementary Fig. 4\textbf{f} & 125 nA & 325 nA & 450 nA & 242 $\Omega$ & 110 $\Omega$ & 110 $\Omega$ \\
    \hline
\end{tabular}
\caption{Table of parameter values used to produce plots in Supplementary Fig. \ref{sgating_0102}}
\label{t1}
\end{table}

\section{Conductance data from Device 2 and Device 3}
Conductance data for Device 2 and Device 3 showing accessibility of single mode regime coexistent with superconductivity in all three legs of the devices are shown in Supplementary Fig. \ref{qpc_data_dev2} \textbf{a,b,c} and Supplementary Fig. \ref{qpc_data_dev3} \textbf{a,b,c}, respectively. For Device 3 we also observe structures similar to Coulomb diamonds for high negative gate voltage.  Data measured at elevated temperature of 2.1 K and at out-of plane magnetic field of $B=0.95\rm{T}$ for the Device 2 smooths out conductance data (Supplymentry Fig. \ref{qpc_data_dev2} e, f), due to suppression of coherent backscattering.

\begin{figure}[ht]
    \centering
    \includegraphics[width=1\textwidth]{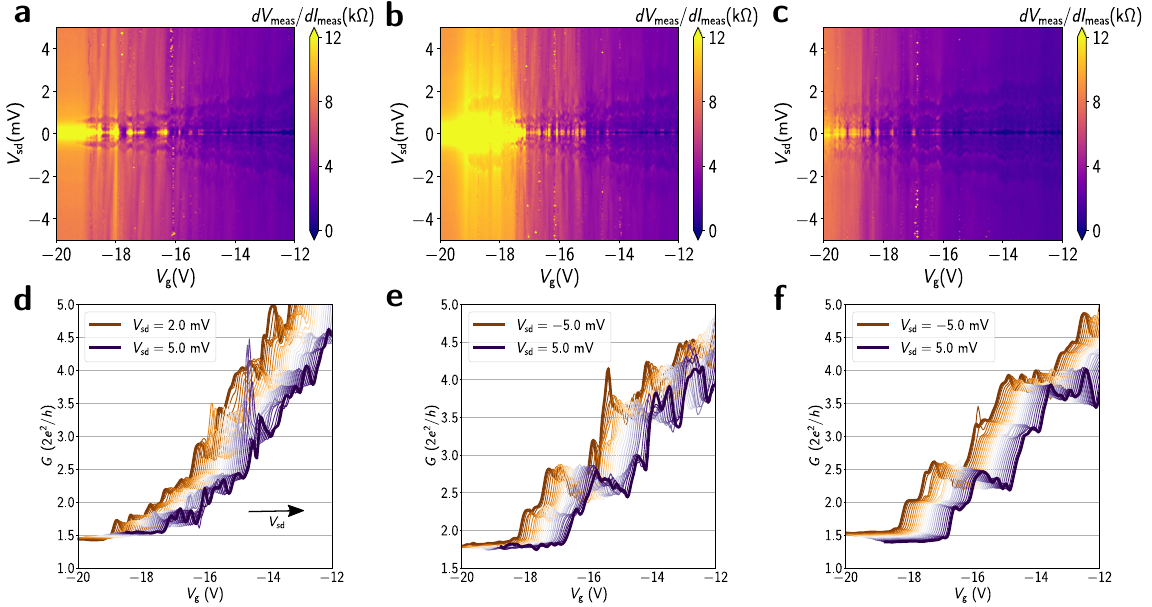}
    \caption{Color map of differential resistance as a function of source-drain bias $\rm{V_{\textrm{sd}}}$ and gate voltage $\rm{V_\textrm{g}}$ for Device 2 at $B=0$ and $T=90$ mK \textbf{a} for terminal pair 0 and 1,  \textbf{b} terminal pair 0 and 2 and \textbf{c} terminal pair 1 and 2. Differential conductance as a function of gate voltage for different $V_{\rm{sd}}$ for Device 2 at \textbf{d} $B=0$ and $T=90$ mK   \textbf{e} at $B=0$ and $T=2$ K \textbf{f} at $B=0.95$ T and $T=90$ mK for terminal pair 0 and 1.  The curves correspond to increments in $V_{\rm{sd}}$ of 0.125 mV, and are offset on the gate voltage (arrow indicating direction of increasing $V_{\rm{sd}}$) by 3 mV for clarity. The $V_{\rm{sd}}$ range is shown in plot legends.
    }
    \label{qpc_data_dev2}
\end{figure}

\begin{figure}[ht]
    \centering
    \includegraphics[width=1\textwidth]{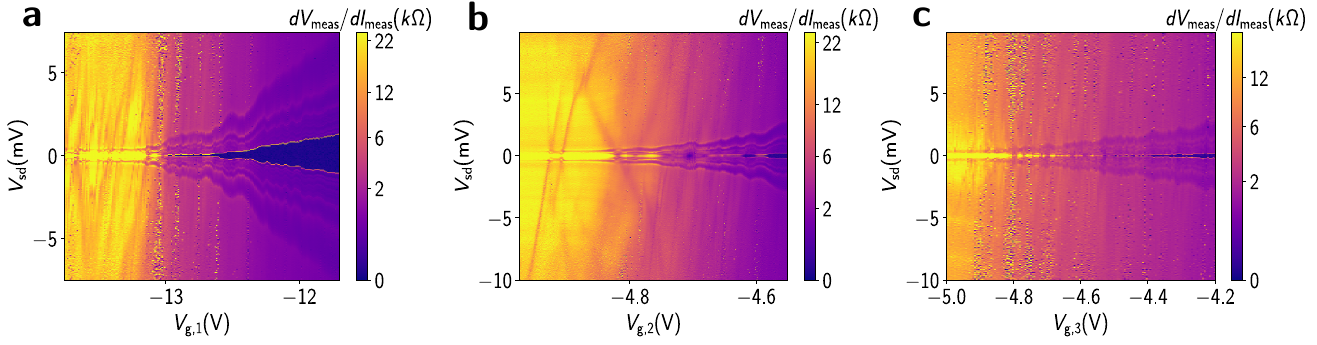}
    \caption{Color map of differential resistance as a function of source-drain bias $\rm{V_{\textrm{sd}}}$ and split gate voltages for Device 3 at $B=0$ and $T=50$ mK for \textbf{a} terminal pair 0 and 1 at $V_{\textrm{g},2}=-5.1$ V and $V_{\textrm{g},3}=-7.5$ V while $V_{\textrm{g},1}$ is swept. \textbf{b} terminal pair 0 and 2 at $V_{\textrm{g},1}=-6$ V and $V_{\textrm{g},3}=-3$ V while $V_{\textrm{g},2}$ is swept and \textbf{c} terminal pair 1 and 2 at $V_{\textrm{g},1}=-7$ V and $V_{\textrm{g},2}=-7$ V while $V_{\textrm{g},3}$ is swept.}
    \label{qpc_data_dev3}
\end{figure}

\FloatBarrier

\section{Data from Device 4}
Here we present data from a fourth Device lithographically identical to Device 1 and Device 3. We see similar transport features, showing a high degree of reproducibility of such devices.
\begin{figure}[ht]
    \centering
    \includegraphics[width=1.0\textwidth]{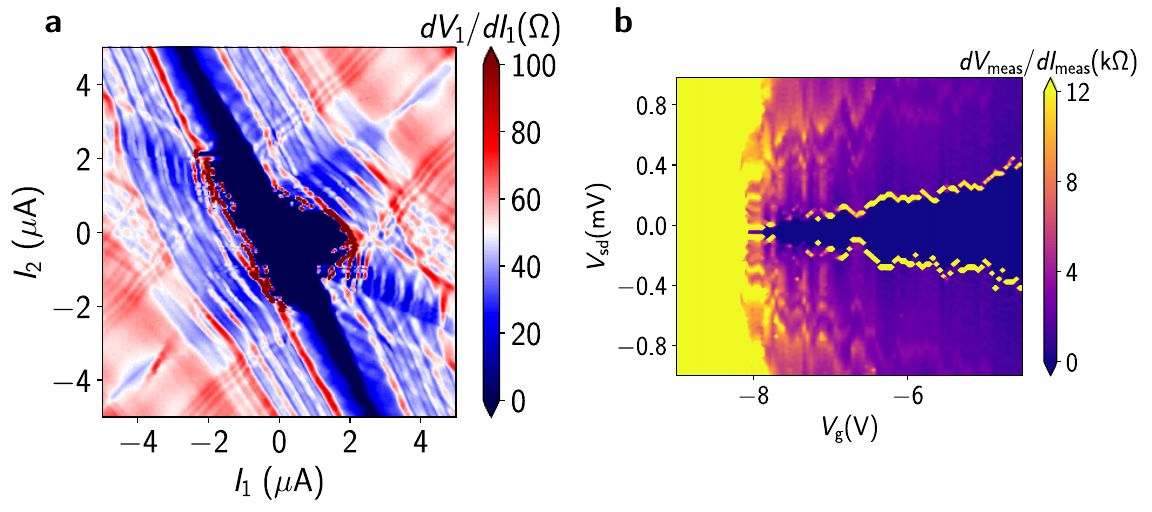}
    \caption{\textbf{a}  Measurement of $dV_1/dI_1$ on Device 4 at small magnetic field and $T=30$ mK. \textbf{b} Differential conductance as a function of gate voltage and $V_{\rm{sd}}$ for Device 4 at small magnetic field and $T=30$ mK}
    \label{dev3}
\end{figure}